\documentclass[12pt]{iopart}

\usepackage{iopams,booktabs}
\usepackage{filecontents}
\usepackage{graphicx,xcolor,marginnote}
\usepackage[utf8x]{inputenc}

\renewcommand\emph[1]{{\color{red}{\sffamily\textit{#1}}}}
\graphicspath{{figures/}} 
\usepackage{caption}
\usepackage{subcaption}
\usepackage{adjustbox}
\usepackage[colorlinks, citecolor = blue, urlcolor = blue]{hyperref}
\newcommand{\cc}{CrCl$_3$}
\newcommand{\vc}{VCl$_3$}
\newcommand{\cb}{CrBr$_3$}
\newcommand{\vb}{VBr$_3$}
\begin{document}
\title{A Study of Electronic and Magnetic Properties of Transition Metal Trihalides}
\author{Shrestha Dutta, Sachin Varma U, Payel Bandyopadhyay and Rudra Banerjee}

\address{Department of Physics and Nanotechnology, SRM Institute of Science and Technology, Kattankulathur, Tamil	Nadu, 603203,
	India}
\ead{bnrj.rudra@gmail.com}


\begin{abstract}
	We present the electronic and magnetic structure calculations of VCl$_3$, VBr$_3$, CrCl$_3$ and CrBr$_3$. The results are
	obtained by density functional theory with plane wave basis sets. The transition metal trihalides generally optimize either
	in trigonal or monoclinic structures. We have focused on the effect of symmetry on the electronic and magnetic properties of
	the systems. We have found that magnetic moments change considerably depending on the symmetry. Both CrX$_3$ has shown a
	bandgap $\approx$ 2eV while the V based systems have shown half-metallic properties.
\end{abstract}

\section{Introduction}
The family of transition metal trihalides MX$_3$,  where M is a transition metal cation (M=Ti, V, Cr, Fe, Mo, Ru, Rh, Ir) and
X is a halogen anion (X=Cl, Br, I), have been known for more than 50 years\cite{Lin_1993,Michel_1971}.
These compounds have been in spotlight for some amusing electronic and magnetic properties they exhibit in their bulk and
monolayer phases such as, intrinsic long-range magnetic order in atomically thin layers\cite{Wang_2022} as well as its
profound applications in spintronics\cite{Basak_2023}.
With lower dimensionalities of MX$_3$, their bandgap can be tuned by doping or changing their stacking orientation.
This property of MX$_3$ opens up the opportunity in multi-purpose applications.

Structurally an X-ray and neutron diffraction study ascertained that the single MX$_3$ crystal adopts a monoclinic AlCl$_3$ structure
(space group $C2/m$), termed as ($\beta$~phase)\cite{Kratochv_lov__2022}. $\beta$-phase is found to be lower symmetry
crystallographic phase which is most likely to be found at higher temperatures. Upon cooling from higher temperature, MX$_3$
layers rearrange themselves from $\beta$-phase to BiI$_3$-type trigonal(space group R$\bar{3}$) structure known as $\alpha$-phase.
The transition has been reported to be a first-order phase transition which does not involve any magnetic changes\cite{Craco_2021}.
This phase transition is believed to result from interlayer interactions among the layers\cite{McGuire_2017,McGuire_2015}.
The interlayer interactions are caused by weak van der Waals(vdW) bonding\cite{McGuire_2017} between halogen(X) anions hence,
known as vdW structures. The vdW structures reveal themselves to be truly captivating in the field of materials science due
to the presence of intrinsic magnetism and magnetic anisotropy\cite{Albaridy_2020}, tunable band gap and high temperature magnetic
ordering\cite{Basak_2023}. This property of tunable band gap leads to new generation spintronics, magnetic and magneto-optic
applications\cite{Tomar_2019}.

Spintronics is a new generation information technology where spins of electrons are employed as information carriers and also
speeds up data processing. Half metals, spin gapless semiconductors and bipolar magnetic semiconductors were proposed one after
another as probable property in spintronics\cite{Li_2016}. Half metals have been established commendatory in spin current
generation and making spintronic and other nano-electronic devices. Dirac half metals exhibit a large bandgap in one spin channel
and metallic character in the other. We attempt to figure out half-metals among the family of transition metals in bulk phases.

For many of the  MX$_3$ compounds downscaling the bulk phase to a stable monolayer is still a challenge.
Some studies suggest that monolayers with binding energy smaller than 0.15 eV per atom might be feasible for
exfoliation\cite{Singh_2015}.
\begin{figure}[htpb]
	\centering
	\begin{subfigure}[b]{0.39\textwidth}
		\includegraphics[width=0.8\textwidth]{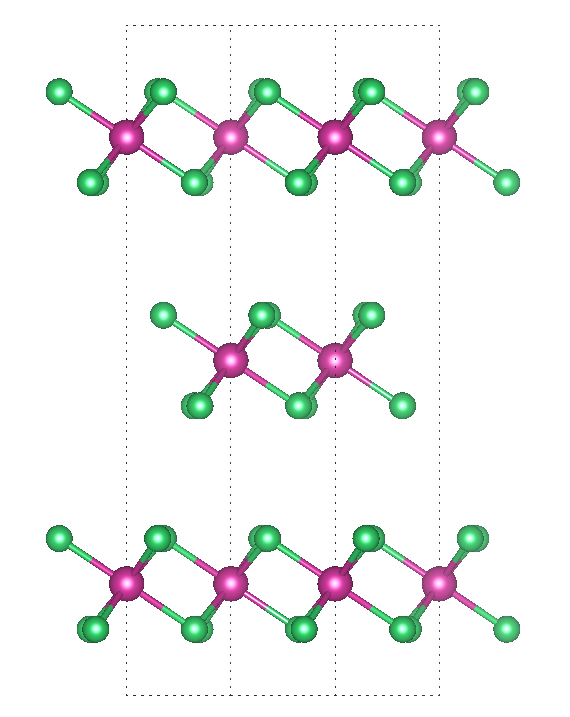}
		\caption{R$\bar{3}$}
		\label{fig:trigonal}
	\end{subfigure}
	\hspace{0.8cm}
	\begin{subfigure}[b]{0.39\textwidth}
		\includegraphics[width=0.8\textwidth]{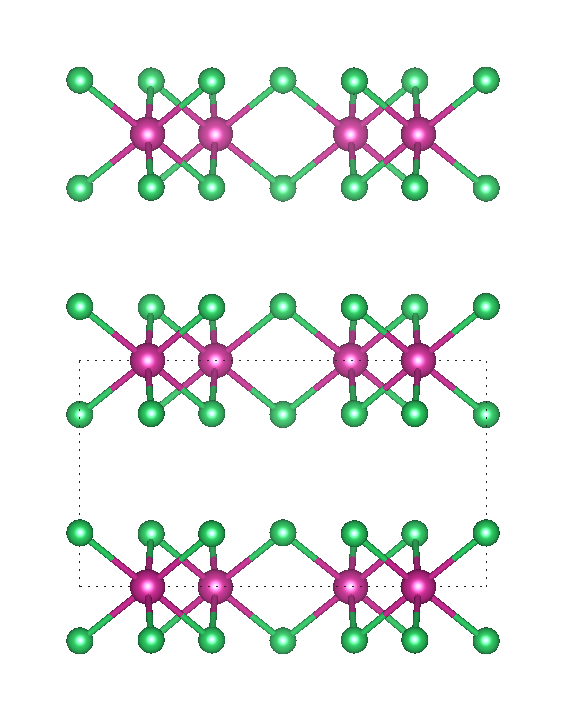}
		\caption{$C2/m$}
		\label{fig:monoclinic}
	\end{subfigure}
	\caption{Bulk structures of MX$_3$ for (\subref{fig:trigonal})$R\bar{3}$ and (\subref{fig:monoclinic})$C2/m$ space groups respectively}
	\label{fig:three_graphs}
\end{figure}
The physical properties of a few-layer or monolayer structures of such transition metal MX$_3$ are quite different from that
of their bulk structures as a result of the different screening environments they experience due to the differences in their
effective dimensionality\cite{Huang_2017}.

Various density functional theory(DFT) based studies of bulk and few-layer MX$_3$ for the
ground-state have been carried out theoretically\cite{Craco_2021,Qin_2021,Pascazio_2021,Kvashnin_2022}.
The difference in electronic screening for bulk and few-layer MX$_3$\cite{Acharya_2022} results in different electronic bandgaps.
These studies cover the interlayer exchange coupling\cite{Soriano_2019} as well as its dependence on the stacking of layers,
possible ways to stabilize skyrmions\cite{Behera_2019}, and the electric-field switching of the magnetization\cite{Su_rez_Morell_2019}.

Researchers have detected rich magnetic ground states in the MX$_3$ family.
Magnetism in the ground state of a material is correlated to interlayer vdW coupling.
While vdW coupling originates from distinct stacking orders and the signature Kagome lattice\cite{Nharangatt_2022}.

In MX$_3$, magnetism originates from the angular momentum associated with partially filled $d$ orbitals.
It has been predicted that the in-plane magnetic interaction
among the transition metal cations is a consequence of super-exchange through shared coordinating halogen anions.
The super-exchange interaction depends upon factors like orbital occupations and the M-X-M angle\cite{khomskii_2014}.

Among the MX$_3$ series, the Cr and V-based materials are mainly of importance. \cc~has recently been shown to have an unusual
magnetic easy axis normal to the $c$-axis\cite{Cai_2019}. Further, Ebrahimian \textit{et al.} has recently shown that the
magnetism in bilayer \cc~can be easily controlled by strain and electric field\cite{Ebrahimian_2023}. Ahmad \textit{et al.} has
shown magnetism of \cc~can be controlled by pressure also\cite{Ahmad_2020}. \vc~has been shown to have half metallicity in
2D\cite{Zhou_2016}. Doping $3d$ transition metal to \vc~has been shown to tune its electronic and magnetic properties very
recently\cite{Ouettar_2023}. High tunnel magnetoresistance and spin filtering effect are shown in \cb\cite{Liu_2021}. Grzeszczyk
\textit{et al.} has recently shown exciton magnetization in \cb\cite{Grzeszczyk_2023}.

From the above discussion, it is clear that V and Cr-based MX$_3$ systems play a significant role in the development of spintronics.
Hence, a detailed and systematic study of these systems from the equivalent theoretical field is necessary.  In this
study, we compare the structural and electronic properties of the possible space groups ($R\bar{3}$ and $C2/m$) of
trichlorides and tribromides of V and Cr using \textit{ab-inito} density functional theory.

\section{Computational Details}
The Density functional theory (DFT) calculations of vdW layered MX$_3$ have been brought about with Quantum
Espresso\cite{Giannozzi_2009,Giannozzi_2017}. For the exchange-correlation functional general gradient approximation (GGA) was
used in the projector augmented wave(PAW)\cite{Kresse_1999} pseudopotential.  We used the plane-wave cut-off energy of 320
eV to treat interactions between the valence electrons and ion cores.  The calculations are done in $\Gamma$-point.

Since MX$_3$ is a layered system, vdW corrections were used based on the semiempirical Grimme’s DFT-D2\cite{Grimme_2010}
approach. We need to account for the strong correlation effects of localized $3d$ electrons of transition metal atoms
to describe their electronic and magnetic properties\cite{Wang_2020} correctly.
In this case, a simplified version of the DFT+U method suggested by Dudarev\cite{Dudarev_1998} was employed.
We set the effective Hubbard interaction parameter U$_{eff}$ of metal atoms to 2.7 eV\cite{Sun_2020,Rubyann_2022}.
It is acquired by comparing physical properties directly.
\section{Results and Discussions}
The MX$_3$ system stabilizes in $R\bar{3}$ and $C2/m$ structures. We have calculated the ground state of both structures with
M=V, Cr and X=Cl, Br.
The optimized lattice parameter(a), optimized M-X bond length(d$_{M-X}$), interplanar separation(d$_0$)and magnetic
moment per formula unit(\textit{f.u.}) for VCl$_3$, VBr$_3$, CrCl$_3$ and CrBr$_3$ are shown in Table (\ref{tab1}), for FM magnetic
ordering. The minimal energy ground state of the Bravis lattice are shown in bold.

\begin{table}[htpb]
	\begin{center}
		\scalebox{0.9}{
			\begin{tabular}{lcccp{.2\textwidth}}
				\toprule
				MX$_3$                         & a(\AA)                               & d$_{M-X}$(\AA) & d$_0$(\AA) & magnetic
				moment($\mu_B$/\textit{f.u.})                                                                                                        \\
				\toprule
				\bfseries{VCl$_3$(R$\bar{3}$)} & 6.12                                 & 2.40           & 4.48
				                               & 2.54(2.0 \cite{mp-28117})                                                                           \\
				VCl$_3$($c2/m$)                & 6.26                                 & 2.40           & 5.08
				                               & 2.32(2.96 \cite{Kratochv_lov__2022})                                                                \\
				\bfseries{VBr$_3$(R$\bar{3}$)} & 6.52                                 & 2.56           & 4.60
				                               & 2.84(2.6 \cite{Kong_2019})                                                                          \\
				VBr$_3$($c2/m$)                & 6.62                                 & 2.57           & 5.15       & 2.46(2.0 \cite{hovančík_2023}) \\
				CrCl$_3$(R$\bar{3}$)           & 6.08                                 & 2.38           & 4.32       & 3.59(3.0 \cite{Soriano_2020})  \\
				\bfseries{CrCl$_3$($c2/m$)}    & 6.09                                 & 2.38           & 5.15       & 5.38
				(6.0\cite{Liu_2016})                                                                                                                 \\
				{CrBr$_3$(R$\bar{3}$)}         & 6.46                                 & 2.55           & 4.75
				                               & 3.91(3.10\cite{Zhang_2015})                                                                         \\
				\bfseries{CrBr$_3$($c2/m$)}    & 6.39                                 & 2.56           & 5.34       & 3.90                           \\\bottomrule
			\end{tabular}
		}
	\end{center}
	\caption{Crystal and magnetic structure information of MX$_3$ compounds. The magnetic moments in the bracket show their
		experimental values.}
	\label{tab1}
\end{table}

The magnetic moments match   closely with their experimental values (given in the bracket and reference therein). The
discrepancies come from the fact that we have done a $\Gamma$-point only calculation.

\begin{figure}[htpb]
	\centering
	\begin{subfigure}[b]{.4\textwidth}
		\begin{center}
			\includegraphics[width=\textwidth]{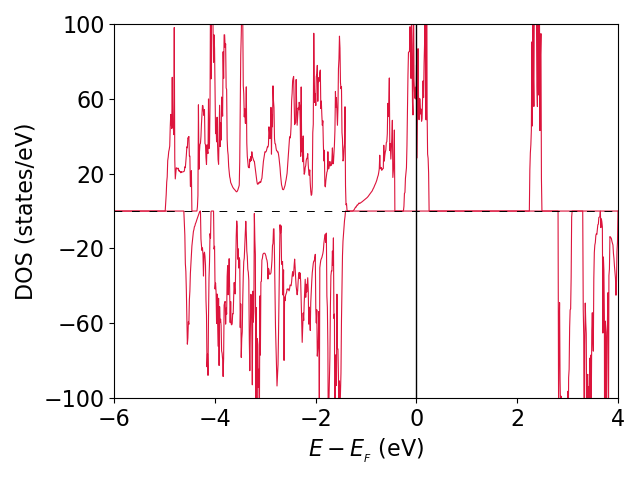}
		\end{center}
		\caption{$R\bar{3}$}
		\label{fig:vcl_r3}
	\end{subfigure}
	\begin{subfigure}[b]{.4\textwidth}
		\begin{center}
			\includegraphics[width=\textwidth]{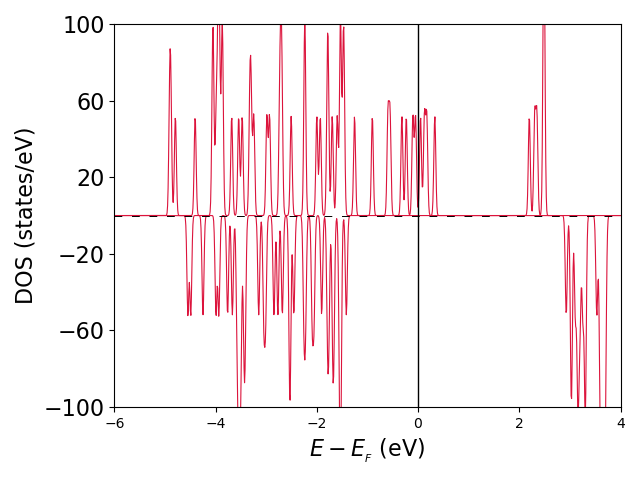}
		\end{center}
		\caption{$c2/m$}
		\label{fig:vcl_c2m}
	\end{subfigure}
	\caption{DOS of VCl$_3$ for (\subref{fig:vcl_r3}) $R\bar{3}$ and (\subref{fig:vcl_c2m}) $c2/m$ structure }
	\label{fig:vcl3}
\end{figure}
\begin{figure}[htpb]
	\centering
	\begin{center}
		\includegraphics[width=.4\textwidth]{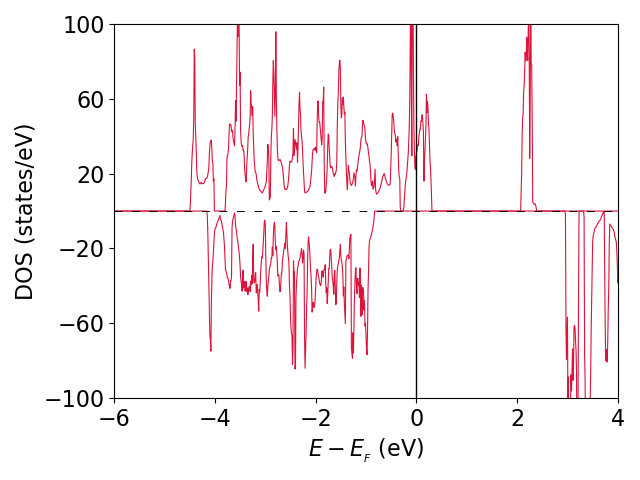}
	\end{center}
	\caption{DOS of VBr$_3$ for $R\bar{3}$}
	\label{fig:VCl3}
\end{figure}
\begin{figure}[htpb]
	\centering
	\begin{subfigure}[b]{.4\textwidth}
		\begin{center}
			\includegraphics[width=\textwidth]{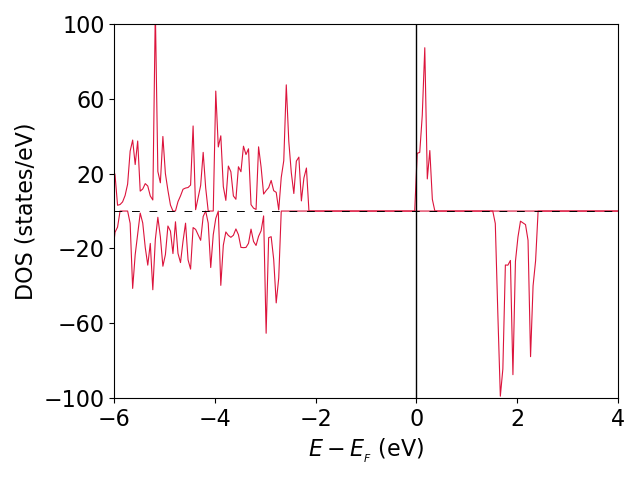}
		\end{center}
		\caption{$R\bar{3}$}
		\label{fig:crcl_r3}
	\end{subfigure}
	\begin{subfigure}[b]{.4\textwidth}
		\begin{center}
			\includegraphics[width=\textwidth]{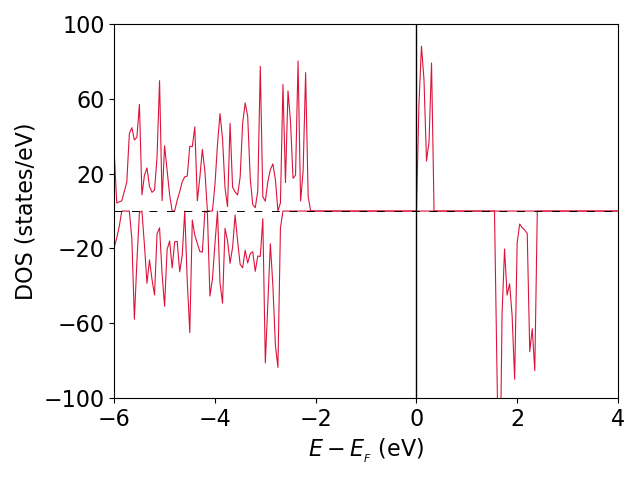}
		\end{center}
		\caption{$c2/m$}
		\label{fig:crcl_c2m}
	\end{subfigure}
	\caption{DOS of CrCl$_3$ for (\subref{fig:crcl_r3}) $R\bar{3}$ and (\subref{fig:crcl_c2m}) $c2/m$ structure }
	\label{fig:crcl3}
\end{figure}
\begin{figure}[htpb]
	\centering
	\begin{subfigure}[b]{.4\textwidth}
		\begin{center}
			\includegraphics[width=\textwidth]{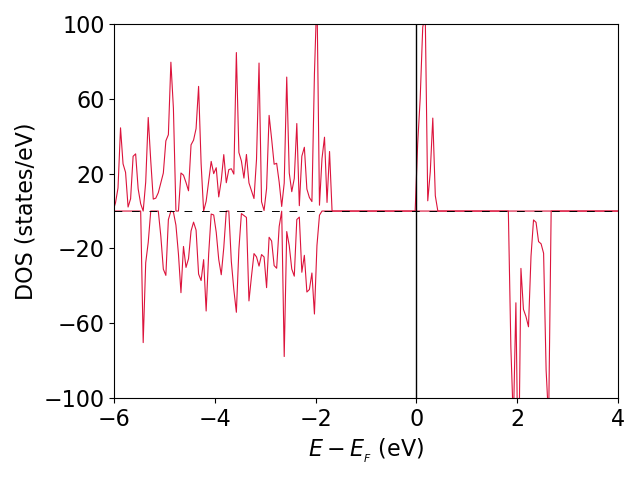}
		\end{center}
		\caption{$R\bar{3}$}
		\label{fig:crbr_r3}
	\end{subfigure}
	\begin{subfigure}[b]{.4\textwidth}
		\begin{center}
			\includegraphics[width=\textwidth]{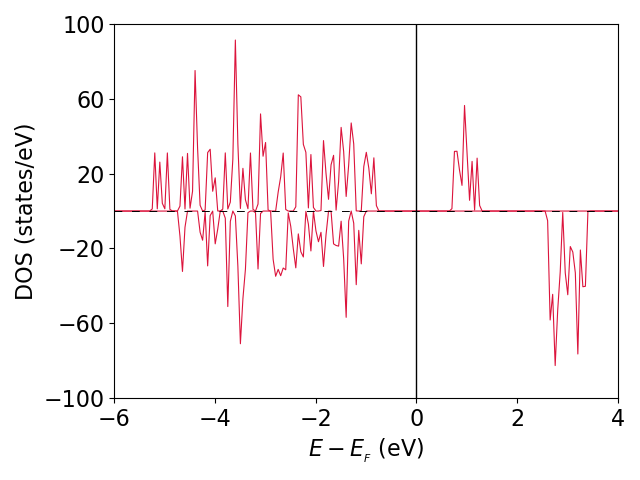}
		\end{center}
		\caption{$c2/m$}
		\label{fig:crbr_c2m}
	\end{subfigure}
	\caption{DOS of CrBr$_3$ for (\subref{fig:crbr_r3}) $R\bar{3}$ and (\subref{fig:crbr_c2m}) $c2/m$ structure }
	\label{fig:crbr3}
\end{figure}

In figures (\ref{fig:vcl3})-(\ref{fig:crbr3}), we have discussed the electronic structure of MX$_3$ systems.

The DOS of VCl$_3$ shows prominent states in the majority spin channels and a big gap in minority spin channel for both
$R\bar{3}$(Fig. (\ref{fig:vcl_r3})) and $c2/m$(Fig. (\ref{fig:vcl_c2m})) structure. This resembles the
experimental and previous theoretical findings\cite{Tomar_2019}.

VBr$_3$ DOS also shows the same characteristics. These half-metallic behavior of VCl$_3$ and VBr$_3$ is well known and sustained
in bilayer systems\cite{Liu_2020}.

With our very small U$_{eff}$, the story is completely different for Cr-based trihalids. Both CrCl$_3$ and CrBr$_{3}$ show bandgap
$\approx 2$eV in all the cases. This is also well known feature\cite{Wang_2011}. Though our position of DOS peaks does not
matche the existing literature, the characteristics are well produced with the small $U_{eff}$. The band gap decreases
significantly from CrCl$_3$ to CrBr$_3$.

\section{Conclusion}
Although there are several studies on MX$_3$ systems, there is still a gap in understanding their stacking, structure, and
electronic and magnetic properties. Due to the recent interest in spintronic applications, it was necessary to study those
materials systematically. In this study, we have explored the electronic and magnetic properties of \cc, \cb, \vc~and \vb~in bulk
phase using DFT with plane wave basis. We have shown that the stacking and the space group play an important role in their
magnetic and electronic properties. We are confident that this will enable us to design better MX$_3$ based spintronic devices with
bilayer and multilayer.
\section*{References}
\bibliographystyle{iopart-num}
\bibliography{icon}

\providecommand{\newblock}{}
\begin{thebibliography}{10}
\expandafter\ifx\csname url\endcsname\relax
  \def\url#1{{\tt #1}}\fi
\expandafter\ifx\csname urlprefix\endcsname\relax\def\urlprefix{URL }\fi
\providecommand{\eprint}[2][]{\url{#2}}

\bibitem{Lin_1993}
Lin J and Miller G~J 1993 {\em Inorganic Chemistry\/} {\bf 32} 1476--1487
  \urlprefix\url{https://doi.org/10.1021%2Fic00060a025}

\bibitem{Michel_1971}
Michel C, Moreau J~M and James W~J 1971 {\em Acta Crystallographica Section
  B\/} {\bf 27} 501--503
  \urlprefix\url{https://doi.org/10.1107/S0567740871002437}

\bibitem{Wang_2022}
Wang Q~H, Bedoya-Pinto A {\em et~al.\/} 2022 {\em {ACS} Nano\/} {\bf 16}
  6960--7079 \urlprefix\url{https://doi.org/10.1021%2Facsnano.1c09150}

\bibitem{Basak_2023}
Basak K, Ghosh M, Chowdhury S and Jana D 2023 {\em Journal of Physics:
  Condensed Matter\/} {\bf 35} 233001
  \urlprefix\url{https://doi.org/10.1088%2F1361-648x%2Facbffb}

\bibitem{Kratochv_lov__2022}
Kratochv{\'{\i}}lov{\'{a}} M, Dole{\v{z}}al P, Hovan{\v{c}}{\'{\i}}k D,
  Posp{\'{\i}}{\v{s}}il J, Bendov{\'{a}} A, Du{\v{s}}ek M, Hol{\'{y}} V and
  Sechovsk{\'{y}} V 2022 {\em Journal of Physics: Condensed Matter\/} {\bf 34}
  294007 \urlprefix\url{https://doi.org/10.1088%2F1361-648x%2Fac6d38}

\bibitem{Craco_2021}
Craco L, Carara S~S, Shao Y~C, Chuang Y~D and Freelon B 2021 {\em Physical
  Review B\/} {\bf 103}
  \urlprefix\url{https://doi.org/10.1103%2Fphysrevb.103.235119}

\bibitem{McGuire_2017}
McGuire M~A, Clark G, KC S, Chance W~M, Jellison G~E, Cooper V~R, Xu X and
  Sales B~C 2017 {\em Physical Review Materials\/} {\bf 1}
  \urlprefix\url{https://doi.org/10.1103%2Fphysrevmaterials.1.014001}

\bibitem{McGuire_2015}
McGuire M~A, Dixit H, Cooper V~R and Sales B~C 2015 {\em Chemistry of
  Materials\/} {\bf 27} 612--620
  \urlprefix\url{https://doi.org/10.1021%2Fcm504242t}

\bibitem{Albaridy_2020}
Albaridy R, Manchon A and Schwingenschl\"{o}gl U 2020 {\em Journal of Physics:
  Condensed Matter\/} {\bf 32} 355702
  \urlprefix\url{https://doi.org/10.1088%2F1361-648x%2Fab8986}

\bibitem{Tomar_2019}
Tomar S, Ghosh B, Mardanya S, Rastogi P, Bhadoria B, Chauhan Y~S, Agarwal A and
  Bhowmick S 2019 {\em Journal of Magnetism and Magnetic Materials\/} {\bf 489}
  165384 \urlprefix\url{https://doi.org/10.1016%2Fj.jmmm.2019.165384}

\bibitem{Li_2016}
Li X and Yang J 2016 {\em National Science Review\/} {\bf 3} 365--381
  \urlprefix\url{https://doi.org/10.1093%2Fnsr%2Fnww026}

\bibitem{Singh_2015}
Singh A~K, Mathew K, Zhuang H~L and Hennig R~G 2015 {\em The Journal of
  Physical Chemistry Letters\/} {\bf 6} 1087--1098
  \urlprefix\url{https://doi.org/10.1021%2Fjz502646d}

\bibitem{Huang_2017}
Huang B, Clark G {\em et~al.\/} 2017 {\em Nature\/} {\bf 546} 270--273
  \urlprefix\url{https://doi.org/10.1038%2Fnature22391}

\bibitem{Qin_2021}
Qin G, Li S and Shao X 2021 {\em AIP Advances\/} {\bf 11}

\bibitem{Pascazio_2021}
Pascazio R, Zito J and Infante I 2021 {\em {CHIMIA}\/} {\bf 75} 427
  \urlprefix\url{https://doi.org/10.2533%2Fchimia.2021.427}

\bibitem{Kvashnin_2022}
Kvashnin Y~O, Rudenko A~N, Thunstr\"{o}m P, Rösner M and Katsnelson M~I 2022
  {\em Physical Review B\/} {\bf 105}
  \urlprefix\url{https://doi.org/10.1103%2Fphysrevb.105.205124}

\bibitem{Acharya_2022}
Acharya S, Pashov D, Rudenko A~N, R\"{o}sner M, van Schilfgaarde M and
  Katsnelson M~I 2022 {\em npj 2D Materials and Applications\/} {\bf 6}
  \urlprefix\url{https://doi.org/10.1038%2Fs41699-022-00307-7}

\bibitem{Soriano_2019}
Soriano D, Cardoso C and Fern{\'{a}}ndez-Rossier J 2019 {\em Solid State
  Communications\/} {\bf 299} 113662
  \urlprefix\url{https://doi.org/10.1016%2Fj.ssc.2019.113662}

\bibitem{Behera_2019}
Behera A~K, Chowdhury S and Das S~R 2019 {\em Applied Physics Letters\/} {\bf
  114} 232402 \urlprefix\url{https://doi.org/10.1063%2F1.5096782}

\bibitem{Su_rez_Morell_2019}
Morell E~S, Le{\'{o}}n A, Miwa R~H and Vargas P 2019 {\em 2D Materials\/} {\bf
  6} 025020 \urlprefix\url{https://doi.org/10.1088%2F2053-1583%2Fab04fb}

\bibitem{Nharangatt_2022}
Nharangatt B and Chatanathodi R 2022 {\em Journal of Magnetism and Magnetic
  Materials\/} {\bf 564} 170105
  \urlprefix\url{https://doi.org/10.1016%2Fj.jmmm.2022.170105}

\bibitem{khomskii_2014}
Khomskii D~I 2014 {\em Transition Metal Compounds\/} (Cambridge University
  Press) \urlprefix\url{https://doi.org/10.1017%2Fcbo9781139096782}

\bibitem{Cai_2019}
Cai X, Song T {\em et~al.\/} 2019 {\em Nano Letters\/} {\bf 19} 3993--3998
  \urlprefix\url{https://doi.org/10.1021%2Facs.nanolett.9b01317}

\bibitem{Ebrahimian_2023}
Ebrahimian A, Dyrda{\l} A and Qaiumzadeh A 2023 {\em Scientific Reports\/} {\bf
  13} \urlprefix\url{https://doi.org/10.1038%2Fs41598-023-32598-1}

\bibitem{Ahmad_2020}
Ahmad A~S, Liang Y {\em et~al.\/} 2020 {\em Nanoscale\/} {\bf 12} 22935--22944
  \urlprefix\url{https://doi.org/10.1039%2Fd0nr04325g}

\bibitem{Zhou_2016}
Zhou Y, Lu H, Zu X and Gao F 2016 {\em Scientific Reports\/} {\bf 6}
  \urlprefix\url{https://doi.org/10.1038%2Fsrep19407}

\bibitem{Ouettar_2023}
Ouettar C, Yahi H, Zanat K and Chibani H 2023 {\em Physica Scripta\/} {\bf 98}
  025814 \urlprefix\url{https://doi.org/10.1088%2F1402-4896%2Facb093}

\bibitem{Liu_2021}
Liu L, Ye S, He J, Huang Q, Wang H and Chang S 2021 {\em Semiconductor Science
  and Technology\/} {\bf 37} 015006
  \urlprefix\url{https://doi.org/10.1088%2F1361-6641%2Fac3639}

\bibitem{Grzeszczyk_2023}
Grzeszczyk M, Acharya S {\em et~al.\/} 2023 {\em Advanced Materials\/} {\bf 35}
  \urlprefix\url{https://doi.org/10.1002%2Fadma.202209513}

\bibitem{Giannozzi_2009}
Giannozzi P, Baroni S {\em et~al.\/} 2009 {\em Journal of Physics: Condensed
  Matter\/} {\bf 21} 395502
  \urlprefix\url{https://doi.org/10.1088%2F0953-8984%2F21%2F39%2F395502}

\bibitem{Giannozzi_2017}
Giannozzi P, Andreussi O {\em et~al.\/} 2017 {\em Journal of Physics: Condensed
  Matter\/} {\bf 29} 465901
  \urlprefix\url{https://doi.org/10.1088%2F1361-648x%2Faa8f79}

\bibitem{Kresse_1999}
Kresse G and Joubert D 1999 {\em Physical Review B\/} {\bf 59} 1758--1775
  \urlprefix\url{https://doi.org/10.1103%2Fphysrevb.59.1758}

\bibitem{Grimme_2010}
Grimme S, Antony J, Ehrlich S and Krieg H 2010 {\em The Journal of Chemical
  Physics\/} {\bf 132} 154104
  \urlprefix\url{https://doi.org/10.1063%2F1.3382344}

\bibitem{Wang_2020}
Wang Z, Qu S, Xiang H, He Z and Shen J 2020 {\em Materials\/} {\bf 13} 1805
  \urlprefix\url{https://doi.org/10.3390%2Fma13081805}

\bibitem{Dudarev_1998}
Dudarev S~L, Botton G~A, Savrasov S~Y, Humphreys C~J and Sutton A~P 1998 {\em
  Physical Review B\/} {\bf 57} 1505--1509
  \urlprefix\url{https://doi.org/10.1103%2Fphysrevb.57.1505}

\bibitem{Sun_2020}
Sun J, Zhong X, Cui W, Shi J, Hao J, Xu M and Li Y 2020 {\em Phys. Chem. Chem.
  Phys.\/} {\bf 22} 2429--2436
  \urlprefix\url{http://dx.doi.org/10.1039/C9CP05084A}

\bibitem{Rubyann_2022}
Olmos R, Alam S, Chang P~H, Gandha K, Nlebedim I~C, Cole A, Tafti F, Zope R~R
  and Singamaneni S~R 2022 {\em Journal of Alloys and Compounds\/} {\bf 911}
  165034
  \urlprefix\url{https://www.sciencedirect.com/science/article/pii/S0925838822014256}

\bibitem{mp-28117}
Materials {P}roject
  \url{https://next-gen.materialsproject.org/materials/mp-28117} {A}ccessed: 20
  June, 2023

\bibitem{Kong_2019}
Kong T, Guo S, Ni D and Cava R~J 2019 {\em Phys. Rev. Mater.\/} {\bf 3} 084419
  \urlprefix\url{https://link.aps.org/doi/10.1103/PhysRevMaterials.3.084419}

\bibitem{hovančík_2023}
Hovančík D, Kratochvílová M, Haidamak T, Doležal P, Carva K, Bendová A,
  Prokleška J, Proschek P, Míšek M, Gorbunov D~I, Kotek J, Sechovský V and
  Pospíšil J 2023 Robust intralayer antiferromagnetism and tricriticality in
  a van der waals compound: Vbr3 case

\bibitem{Soriano_2020}
Soriano D, Katsnelson M~I and Fernández-Rossier J 2020 {\em Nano Letters\/}
  {\bf 20} 6225--6234 pMID: 32787171

\bibitem{Liu_2016}
Liu J, Sun Q, Kawazoe Y and Jena P 2016 {\em Phys. Chem. Chem. Phys.\/} {\bf
  18} 8777--8784 \urlprefix\url{http://dx.doi.org/10.1039/C5CP04835D}

\bibitem{Zhang_2015}
Zhang W~B, Qu Q, Zhu P and Lam C~H 2015 {\em J. Mater. Chem. C\/} {\bf 3}
  12457--12468 \urlprefix\url{http://dx.doi.org/10.1039/C5TC02840J}

\bibitem{Liu_2020}
Liu L, Yang K, Wang G and Wu H 2020 {\em Journal of Materials Chemistry C\/}
  \urlprefix\url{https://doi.org/10.1039%2Fd0tc03962d}

\bibitem{Wang_2011}
Wang H, Eyert V and Schwingenschl\"{o}gl U 2011 {\em Journal of Physics:
  Condensed Matter\/} {\bf 23} 116003
  \urlprefix\url{https://doi.org/10.1088%2F0953-8984%2F23%2F11%2F116003}

\end{thebibliography}
\end{document}